# Focusing and Scanning through Flexible Multimode Fibers without Access to the Distal End

Shamir Rosen[1,ǂ], Doron Gilboa[1,ǂ], Ori Katz[2,3], Yaron Silberberg[1]

*Affiliation:* [1] Department of Physics of Complex Systems, Weizmann Institute of Science, Rehovot, Israel

[2] Institut Langevin, UMR7587 ESPCI ParisTech and CNRS, INSERM ERL U979, 1 Rue Jussieu, 75005 Paris, France

[3] Department of Applied Physics, The Benin School of Engineering and Computer Science, The Hebrew University of Jerusalem, Jerusalem 91904, Israel

[ǂ]these authors contributed equally to this work

**Multimode fibers (MMFs) are attractive ultra-thin replacements for state-of-the-art endoscopes, but the phase randomization in propagation through MMFs poses a major hurdle for imaging and focusing of light. Recently, this challenge has been addressed by pre-measuring the compensation for the fiber's complex input-output modes relations. Unfortunately, the sensitivity of this approach to fiber bending and temperature variations renders it inappropriate for many applications. Here, we demonstrate a truly endoscopic robust method for controlled in-situ focusing and scanning through a flexible uncharacterized MMF, whereby all the instrumentation is situated at the proximal end. We show that in graded-index (GRIN) fibers, light patterns at the proximal end allow retrieving information about the distal light distribution. We utilize these properties and two-photon fluorescence for robust controlled focusing through bended GRIN fibers. Our results carry potential for lensless two-photon micro-endoscopy.**

Ultra-thin endoscopes are highly desired for many applications, and multimode fibers (MMF) are being explored as ultra-thin lensless replacements for the commonly used endoscopes[1,2]. The difficulty with guiding an image or focusing light through a multimode fiber is the phase-velocity dispersion and coupling between the fiber modes. Any image, even a simple focused light beam, launched into the input facet generates a complex speckle pattern at the fiber output (Figure 1 b). To overcome this obstacle, the beam wavefront can be shaped in such a way that it effectively unscrambles this complex pattern, rendering the fiber a satisfactory image guide[3]. All current methods for determining the correcting wavefront require access to both ends of the fiber for pre-calibration[3–11]. This introduces many technical complications for various applications since the calibration is extremely sensitive to slight bending of the fiber[5,12,13] and even to temperature variations. From a practical point of view, it would be highly desirable to be able to determine the wavefront in-situ without accessing the distal fiber end. Here, we demonstrate such a method for focusing ultrashort pulses through a MMF whereby all the instrumentation is situated at the proximal end. Using nonlinear optical feedback in an epidetection geometry, a diffraction-limited focus is formed at the fiber distal end, and its position is deterministically controlled by exploiting the information encoded in the proximally detected back-propagated fluorescence patterns distributions, preserved even in bent GRIN MMF.

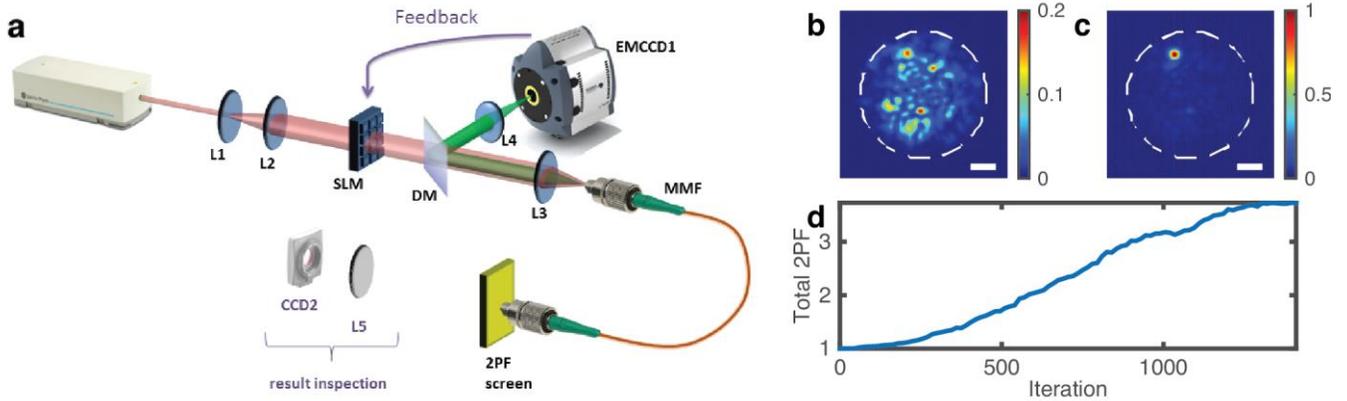

**Figure 1| Experimental Setup.** (a) 100-fs pulses from a Ti:sapphire oscillator are shaped by a spatial light modulator (SLM) and then transmitted through a graded index multimode fiber ( NA=0.2, 50μm core diameter) with a two-photon fluorescence (2PF) screen at its distal end. The fluorescence is collected at the proximal end via the same fiber, using a dichroic mirror (DM) and an EMCCD. The camera is imaging the Fourier plane of the fiber input facet. The proximally detected fluorescence image is used as the feedback signal for controlling the input wavefront with the SLM. An additional camera (CCD2) is used only for inspecting the resulting intensity pattern at the distal end. Images of the intensity pattern at the distal end before (b) and after (c) an optimization of the total proximally detected 2PF, demonstrating focusing to a single speckle grain. The white dashed circle represents the borders of the fiber output facet. Scale bars are 10 microns. (d) The value of the total proximally detected 2PF during the optimization process.

Many recent advancements in utilizing MMFs as ultrathin endoscopes are based on digital wavefront shaping[14,15]. The main approaches for determining the optimal correcting wavefront include measuring the transmission matrix[6,10,13], phase conjugation[8,16] and iterative adaptive algorithms[7,15]. Unfortunately, all of these approaches require access to both ends of the fiber during the calibration procedure, which should be repeated after each slight movement or bending of the fiber. Focusing light through a flexible fiber in a truly endoscopic fashion, i.e. without having to access its distal end, requires a feedback signal that can be detected at the proximal end and is indicative of the distal peak intensity and focus position. Here we show that a spatially distributed two-photon fluorescence (2PF), such as the one present in fluorescently tagged samples[1,2,17], together with information encoded in the back-propagated fluorescence patterns in GRIN MMFs possess all of these required features.

**Principle**

We consider a two-photon lensless MMF-based endoscope, where an ultrashort pulse is delivered to a fluorescently tagged sample through the fiber (Fig.1a). The 2PF excited at the distal end by the spatially and temporally distorted pulse is collected by the same fiber and detected at its proximal input end by an EMCCD camera. It is then used as feedback for a wavefront-shaping optimization algorithm, controlling a spatial light modulator (SLM) at the proximal fiber end. The most straightforward approach for optimizing the input wavefront is to maximize the total integrated nonlinear fluorescent signal, which, for an N-photon process, scales with the N-th power of the exciting field intensity[17]. For a thin fluorescent sample, the total detected nonlinear fluorescence power would be:

$$I_{\det} \propto \int \sigma(x,y) I_{out}^{N}(x,y,t) dxdydt$$

where $I_{out}^{N}(x,y,t)$ is the instantaneous optical intensity at the distal end, at each spatial position (x,y) and time (t), and σ(x,y) is the fluorophore absorption distribution. For a linear optical process (N=1) and a homogeneous fluorescent screen, the integrated intensity is simply the total launched energy, and does not contain information about the output field spatial distribution. However, by optimizing the total intensity from a *higher order nonlinear process* (N>1), such as 2PF, a focused beam enhancing a single speckle grain intensity would be formed and the pulse would be focused to a diffraction limited spot (Figure 1 b-c). Most importantly, as we show below, the position of the formed focus can be deterministically determined, controlled and scanned utilizing information embedded in the spatial distribution of the back-propagated 2PF at the proximal end (Figs.2-3).

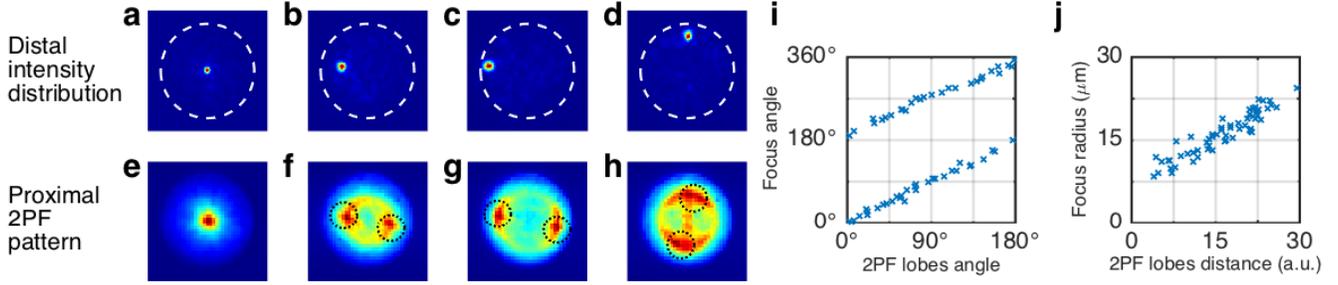

**Figure 2|Images of the focus after optimization taken by CCD2 at the distal end of the fiber (a-d) are shown above the corresponding images of the back-propagated fluorescence taken by CCD1 (e-f). In panels b-d, the focus position was not chosen randomly by the genetic algorithm (GA), but rather set deterministically by a mask (Supplementary Figure 1). Black dotted circles in images (f-h) mark the areas which were set by a mask to contribute positively to the cost function of the GA. The correspondence between the fluorescence patterns and the focus position is plotted both in angle (i) and in radius (j). It can be seen from the angle plot that every fluorescence pattern corresponds to two possible focus positions.**

**Results**

The experimental setup for demonstrating pulse focusing via proximally detected 2PF is depicted schematically in Figure 1(a). ~100fs long pulses at 810nm wavelength are shaped by a phase-only SLM and launched through a 15cm long graded index MMF (NA=0.2, 50μm core diameter, supporting approximately 200 spatial modes at 810nm at each polarization). The pulses excite 2PF from a 50μm-thick 2PF screen placed against the fiber distal end. The back-propagated 2PF that is collected by the same fiber is separated from the excitation light at the proximal end by a dichroic mirror, and the Fourier-transformed image of the fiber facet is recorded by an EMCCD camera. For a unshaped input wavefront a speckled intensity pattern is formed at the fiber distal end (Fig.1b). However, after optimizing the SLM phase pattern to maximize the total 2PF using a genetic algorithm (GA)[18,19], the pulses are tightly focused at the fiber distal end (Fig.1c).

Although this focusing process is very robust and reproducible, yielding intensity enhancements of $\eta \approx 50$ relative to the initial background (Fig.1c), the position of the formed focus with this simple approach is not predetermined. Its position depends on the initial speckle pattern and optimization algorithm. However, as we explain and demonstrate below, the nature of light propagation in a GRIN MMF allows for scanning and controlling the focus position.

Interestingly, due to the self-imaging property of graded index fibers, as the excited fluorescence propagates back through the fiber, information regarding the position of the focus is not entirely lost. Figure 2 shows images of the focused intensity on the 2PF screen at the *distal* end after four different runs of the optimization with different feedback metric, and the corresponding fluorescent patterns as detected at the *proximal* side (Fig.2 e-h). The fluorescent patterns feature a pair of bright lobes at a distance and azimuth, which correlate to the position of the distal focus. In figure 2 (and supplementary movie 1) we plot the correspondence in both radius (Fig. 2 i) and orientation (Fig. 2 j), showing that, *up to a constant rotation angle, one has all the information required to focus at any arbitrary location*. We have utilized these correlations as a precise feedback signal for deterministically controlling the position of the formed focus. Rather than simply maximizing the total 2PF, the different foci positions of Fig.2b-d were obtained by using a spatially modified cost function for the optimization feedback. The cost function was created by multiplying the detected fluorescence image of the fiber proximal facet by a weighting mask set to match the fluorescence lobes corresponding to desired foci positions (dotted circles in Figure f-h, Supplementary Figure 1). Importantly, these correlations are not corrupted by bending of the fiber with radii of curvature of ~5cm.

**Discussion**

To understand the mechanism that leads to the formation of these patterns, we have numerically investigated the expected fluorescence patterns for various inputs and fiber lengths. We found that while in step-index fibers the back-propagating fluorescence is practically evenly distributed over the entire core, in GRIN fibers the transmitted fluorescence indeed forms informative patterns. For a point source input, the self-imaging properties of the GRIN fiber create, periodically, a series of images and mirrored-images as it propagates along the fiber (Supplementary Figures 2-3). While the quality of these images deteriorates with length, they are localized enough to retrieve the input point location even after a surprising long propagation. The periodicity of the imaging sensitively depends on the point source location (see Supplementary movies 2 and 3). The two lobes of fluorescence that are experimentally observed at the proximal end (Fig.2f-h) seem to be formed by the incoherent addition of the images and mirror images of many point sources in the small fluorescent volume excited by a single speckle at the distal end of the fiber. Since there are two symmetric distal focus locations that correspond to nearly identical proximal fluorescence patterns, to unambiguously retrieve the focus position would require using only half the fiber facet. This can be done by creating a permanent obstruction on half of the fiber output facet.

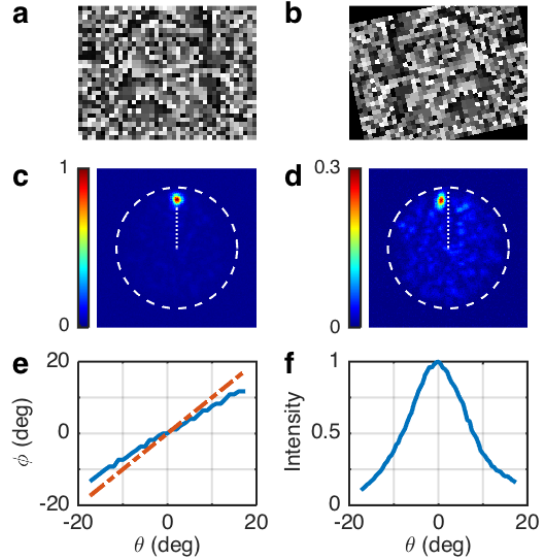

**Figure 3|** Panel (a) shows the optimal SLM pattern found by the GA to enhance a single speckle at a random location (the optimized speckle is shown in c). Rotating the SLM pattern by $12^0$ (b) induces a rotation of $9.2^0$ in the position of the focus (d). Plot (e) shows the dependence of the induced rotation angle ($\phi$) on the rotation angle of the SLM pattern ($\theta$) and plot (f) shows the reduction in peak intensity with the SLM rotation angle ($\theta$).

We also discovered that rotations of the optimized SLM pattern around the fiber axis can directly control, to some extent, the azimuthal position of the focus, due to the partial cylindrical symmetry of the fiber[6]. Figure 3 shows that rotation of the optimized SLM pattern of over ~10 degrees is possible before an increasing loss of wavefront correlation reduce the focus intensity. This allows us to learn a reduced number of focus positions via optimization, and then scan between them.

An additional important challenge in employing MMFs for ultrashort pulses delivery are the temporal distortions induced by both material and modal dispersion[9,20,21]. However, we have previously demonstrated that optimizing a nonlinear signal leads to simultaneous spatial and temporal focusing through a complex scattering medium[19]. In order to investigate to what extent the method is effective in our system, we measured the spatially resolved interferometric autocorrelation of the pulses at the distal end before and after the optimization procedure. While the effect was not strong in the 15cm long graded index fiber, which is designed to have low modal dispersion, a noticeable temporal compression was observed with step index fibers (Supplementary Figure 4).

For in-vivo applications, several limitations of our proof-of-concept experimental system should be addressed, most importantly the wavefront shaping speed. The liquid-crystal based SLM we used here has a ~10Hz refresh rate, which resulted in a duration of the optimization process of the order of 30 minutes. Given sufficiently bright nonlinear signal, using a MEMS-based SLM can reduce this time by more than three orders of magnitude[5]

The ability to focus light selectively on fluorescent targets in a minimally invasive fashion could, in principle, allow various medical applications such as targeting and burning of 2PF tagged targets without harming the surrounding environment. This does not require any active steering of the focus; the beam will 'automatically' focus on the fluorescent tags and will move on to new ones once the old ones are destroyed and photo-bleached.

In another alternative, rather than using 2PF in the sample, fluorescent or second-harmonic[22] coating can be applied to the fiber tip to produce a thin nonlinear medium for the focusing procedure, after which various types of scanning microscopy could be performed. To avoid ambiguity in the retrieval of the focus position, the two halves of the facet can be coated in different colored nonlinear probes.

In this work we have demonstrated focusing of light through a MMF with no pre-calibration and no localized feedback from the distal end of the fiber. We have shown control of the focus position by exploiting the partial cylindrical symmetry of the fiber, and by using the information encoded in the back-propagating fluorescence. In addition, we have shown that in step index fibers, spatial focusing is accompanied by slight temporal compression.

## Methods

**Experimental system.** The experimental setup is presented in Fig. 1. We use a Spectra-Physics Tsunami laser, which produces 100 fs pulses at 810 nm with repetition rate of 80 MHz attenuated to 15 mW. The beam is expanded to cover a circle bounded by the dimensions of the SLM (Hamamatsu LCOS-SLM X10468-02) and then coupled using a microscope objective (X10/0.25 NA) to a MMF (Thorlabs GIF50C (NA=0.2, 50μm core diameter), for Fig. 1-3, Thorlabs FG050UGA (NA=0.22, 50μm core diameter) for Supplementary Fig. 4, both 15 cm in length). The 2PF screen is a 50 $\mu m$-thick capillary filled with disodium fluorescein in ethanol, which is positioned immediately at the fiber output facet. The screen is imaged using a microscope objective (X10/0.25 NA) onto a CCD (iDS uEye LE). For Fig. 1-3 we imaged the excitation field at the laser wavelength, and for Supplementary Fig. 4 we used two shortpass filters (670 nm, 600 nm) to filter out the laser wavelength and image the 2PF. At the input side of the fiber, we separate the back-propagated 2PF from the laser excitation field by a dichroic mirror (Semrock FF720-SDi01) and two bandpass filters (Chroma, D525-250), and record it with an EMCCD (Andor iXon3).

**Optimization process.** The SLM is divided into equally sized square segments. We use a genetic optimization algorithm (Matlab Genetic Algorithm toolbox) to find the phase for each segment that will optimize the 2PF recorded by the EMCCD. For Fig. 1 and 3 we optimized for maximal total intensity by using the sum of the image pixels as the objective function. For Fig. 2 b-d and f-h we used a different objective function, which is calculated by multiplying pixelwise the acquired image with a weighting mask, before summing all values. The mask consists of two lobes positioned symmetrically around the center of the mask. We use different angles and distances between the lobes to focus to different positions (black circles in Fig. 2 f-h, Supplementary Fig. 1). We start the optimization with the SLM divided into 300 segments, and after 300 iterations increase the number of segments to 1200 by dividing each segment into 4. We stopped the optimization for Fig. 1 after 1400 iterations, for Fig. 2 after 1170-1650 iterations and for Fig. 3 after 1065 iterations. For Fig. 2 a, e, i-j we focused the light by maximizing the intensity at different positions on the CCD imaging the fiber distal end, and recorded the matching 2PF patterns on the EMCCD.

**Acknowledgements**

The authors thank Yaron Bromberg for useful discussions and help in numerical simulations. Support by the Adg-ERC project "QUAMI" is gratefully acknowledged.

**Author contributions**

All authors have designed the experiment, analyzed the data and wrote the manuscript. SR and DG have performed the experiments.